# Tuning Ambipolarity in a Polymer Field Effect Transistor using Graphene electrodes

*Kaushik Bairagi [a], Sara Catalano [a], Francesco Calavalle [a], Elisabetta Zuccatti [a], Roger Llopis [a], Fèlix Casanova [a,b], Luis E. Hueso [a,b]\**

[a] CIC nanoGUNE BRTA, 20018 Donostia-San Sebastian, Basque Country, Spain

[b] IKERBASQUE, Basque Foundation for Science, 48013 Bilbao, Basque Country, Spain

*E-mail: l.hueso@nanogune.eu

**Abstract:**

Polymer field-effect transistors with 2D graphene electrodes are devices that merge the best of two worlds: on the one hand, the low-cost and processability of organic materials and, on the other hand, the chemical robustness, extreme thinness and flexibility of graphene. Here, we demonstrate the tuning of the ambipolar nature of the semiconductor polymer N2200 from Polyera ActiveInk™ by incorporating graphene electrodes in a transistor geometry. Our devices show a balanced ambipolar behavior with high current ON-OFF ratio and charge carrier mobilities. These effects are caused by both the effective energy barrier modulation and by the weak electric field screening effect at the graphene-polymer interface. Our results provide a strategy to integrate 2D graphene electrodes in ambipolar transistors in order to improve and modulate their characteristics, paving the way for the design of novel organic electronic devices.

Solution processed polymeric organic semiconductors (OSCs) have become key materials for the development of large area electronics, as they offer mechanical flexibility, low production costs and environmentally friendly applications [1–5]. One of the most challenging steps in the realization of high performance polymer field-effect transistors (FETs) is the choice of the electrode material, as it determines the energy barrier for the charge carrier injection into the semiconductor polymer [6–8]. Noble metals (such as Au, Ag) are one of the obvious choices as source (S) and drain (D) electrodes due to their excellent chemical stability in ambient conditions and suitable work function [1,9,10]. However, due to their large density of state (DOS), the work function of metals is constant and it cannot be changed by a gate voltage bias [11,12]. In this respect, and with the objective of improving the performance of the FETs, it would be ideal to have a gate tunable electrode material so its work function can be adapted to any semiconducting channel [4,13–16].

Graphene fulfills the criteria for being an ideal electrode candidate for polymer electronic devices. Firstly, it is chemically very stable and it can be easily grown on large area using chemical vapor deposition (CVD) [7,14,17–20]. Secondly, its two-dimensional nature produces a very weak screening of the gate electric field. Thirdly, and perhaps more importantly, due to its unique band structure and low DOS, graphene's Fermi level ($E_F$) can be modified with the application of a gate voltage bias [21]. These properties allow to effectively modulate both the energy barrier formed between graphene and a semiconductor as well as the amount of induced charges into such semiconductor [14,22], promoting the performance of diverse hybrid organic electronic devices [4,15,17,21,22].

Ambipolar transistors enable both p- and n-type operations within a single channel material [23–26]. In an ambipolar organic FET, one of the factors that determines its efficient operation is the energy level alignment of the highest occupied molecular orbital (HOMO) and lowest unoccupied molecular orbital (LUMO) of the organic semiconductor with respect to the Fermi level ($E_F$) of the electrode [8,27]. As outlined before, we can take advantage of the properties of graphene electrodes to explore the ambipolar operation of FETs.

As a particular example, we chose Poly{[N,N′-bis(2-octyldodecyl)-naphthalene-1,4,5,8-bis(dicarboximide)-2,6-diyl]-alt-5,5′-(2,2′ bithiophene)} P(NDI2OD-T2) (or N2200, Polyera ActiveInk™) which is a widely used electron transporting polymer [5]. However, recent studies show that this polymer exhibits ambipolarity in the presence of a suitable top dielectric, such as Cyclized Transparent Optical Polymer (CYTOP) ($k$ = 2.1), poly(methyl methacrylate) (PMMA) ($k$ = 3.6), polystyrene (PS) ($k$ = 2.6) or their combinations [5,28,29]. The electron and hole mobilities can also be tuned up to a certain extent using ferroelectric polymers (such as poly[(vinylidenefluoride-co-trifluoroethylene] or P(VDF-TrFE) ) [26]. In all the devices reported previously, metal electrodes were routinely employed [30]. Here, we show the dual gate operation of a solution-processed N2200 transistor with graphene electrodes to study and tune the ambipolar response of the device.

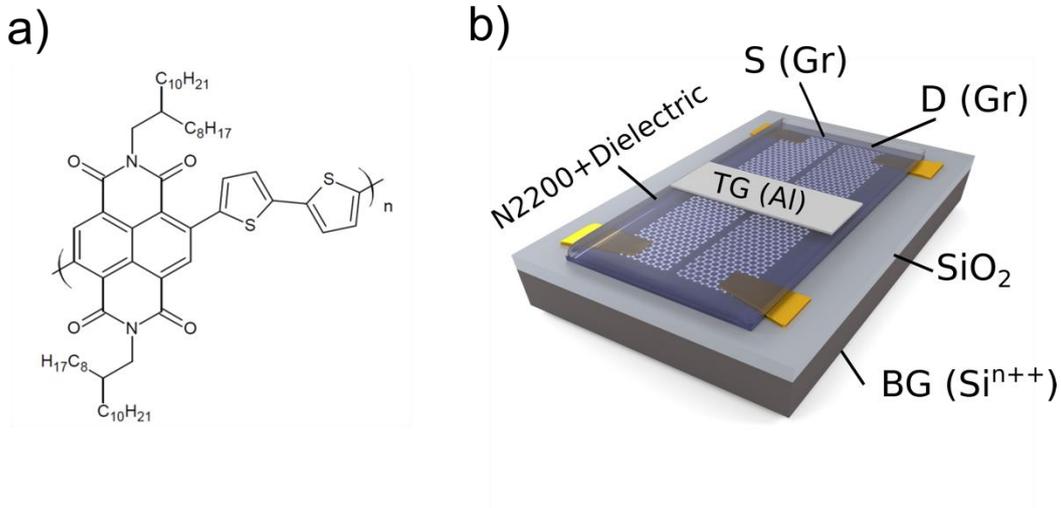

Figure 1: (a) Representation of the molecular structure of N2200. (b) Scheme of the dual gate N2200 transistor architecture with graphene as electrode.

Figure 1a and 1b present the structure of the N2200 polymer and the device architecture, respectively. The device fabrication process is detailed in the experimental section. Figure 2a represents the square resistance ($R_{sq}$) of a typical graphene electrode as a function of the bottom gate bias. We can observe that the graphene is heavily hole-doped with the charge neutrality point (CNP) at 72V, which is common for CVD-grown graphene transferred to a $SiO_2$ substrate [7,14]. Taking into account the capacitance of the bottom gate dielectric (11.5 nF.cm$^{-2}$), one can calculate the induced charge and thereby follow the modulation of the Fermi-energy ($\Delta E_F = E - E_F$) of pristine graphene as a function of the bottom gate voltage (Figure 2b). This calculation is necessary for understanding the energy level alignment of graphene $E_F$ with the HOMO and the LUMO of the N2200 polymer in the rigid band approximation. In the presence of the N2200 polymer, the CNP of the graphene electrode shifts to more positive gate bias, meaning that the charge transfer between graphene and N2200 makes graphene more hole-doped (see Figure S1 for more details).

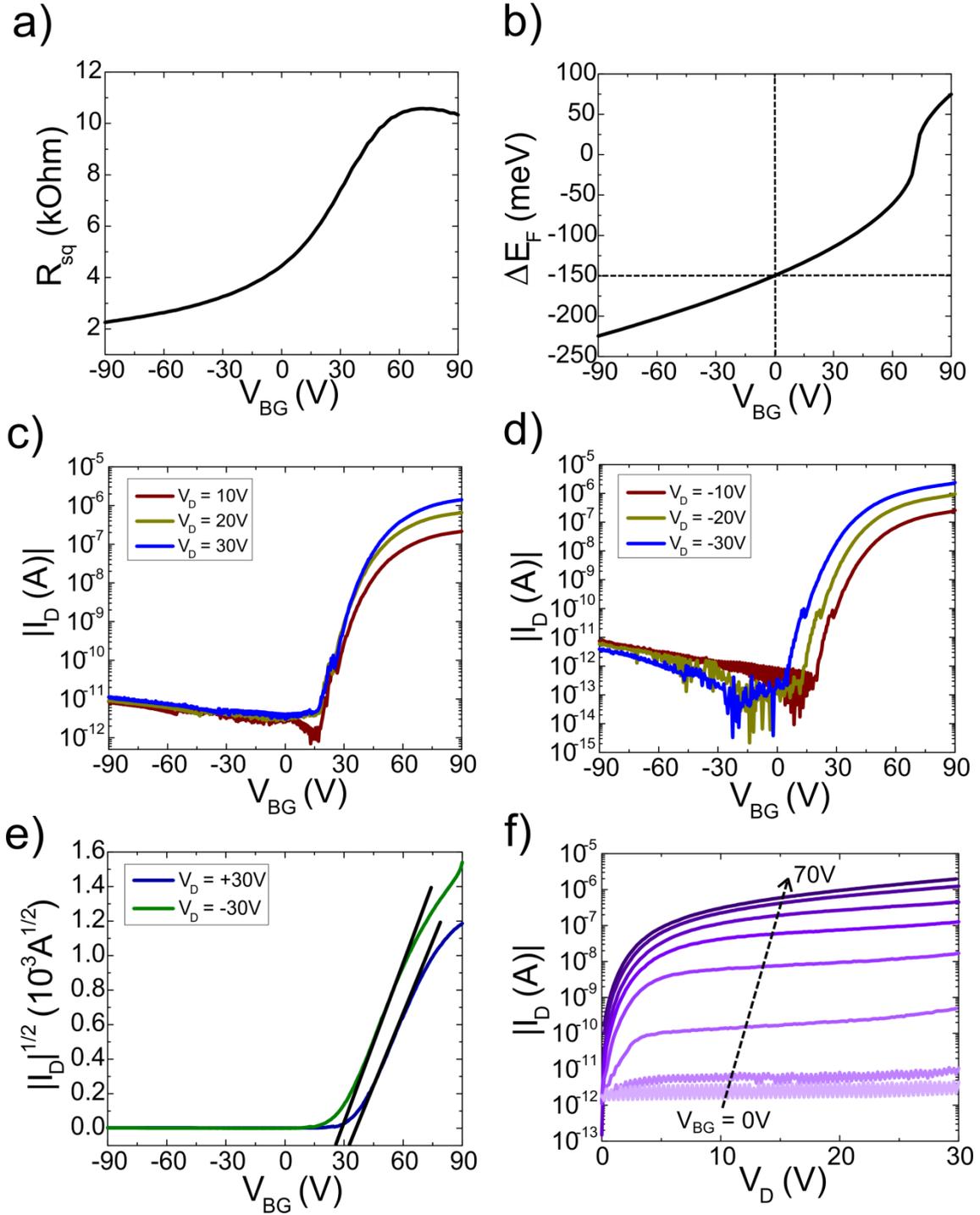

Figure 2: (a) Gate-dependence of the two terminal resistance of a typical graphene electrode. (b) Modulation of the graphene Fermi energy ($\Delta E_F = E - E_F$) as a function of the bottom gate bias. Transfer characteristics of the n-channel operation of N2200 transistor with graphene as source and drain electrodes, (c) with positive $V_D$, (d) with negative $V_D$. (e) Typical $\sqrt{I_D}$ vs. $V_{BG}$ for drain biases of $V_D = +30V$ and $V_D = -30V$; electron mobility is calculated from the slope of the curve. (f) Current-voltage ($I_D - V_D$) output characteristics of the n-channel transistor as a function of bottom gate voltage ($V_{BG}$).

We now focus on the characterization of the bare N2200 lateral FET with graphene electrodes. The transfer curves of the transistor in a bottom gate bottom contact (BGBC) geometry are shown in Figure 2c (for positive drain biases) and Figure 2d (for negative drain biases), respectively. The current ON-OFF ratio reaches up to $10^7$ for negative values of $V_D$. The transfer curves for both positive and negative $V_D$ show no indication of p-type behavior, as expected. The electron mobility values obtained for $V_D = 30V$ and $V_D = -30V$ are $0.8 \times 10^{-3}$ cm$^2$V$^{-1}$s$^{-1}$ and $0.9 \times 10^{-3}$ cm$^2$V$^{-1}$s$^{-1}$, respectively (from Figure 2e), which are higher than those commonly reported with conventional metal electrodes [28]. Typical current-voltage ($I_D - V_D$) characteristics at room temperature as a function of the bottom gate voltage also confirm the good n-type operation of the FET, as shown in Figure 2f. The large current ON-OFF ratio, high electron mobility and optimal n-type operation of the N2200 transistor validate the use of graphene electrodes in these devices [14].

We now move to the specific case in which the N2200 polymer is coated with a top dielectric polymer. Our previous study, with conventional metal electrodes (Ti/Au) in a dual gate FET configuration, has confirmed that ambipolarity can be induced in the bulk of the N2200 polymer in the presence of a top polymer dielectric (PMMA, CYTOP or their combination) [28]. Figure 3a plots the transfer curves of the N2200 FET probed with respect to the bottom gate after PMMA+CYTOP dielectric coating for a drain bias of $V_D$

= +20 V and $V_D$ = -20 V. A clear signature of ambipolarity is observed for both signs of the drain biases, confirming our previous results and independently of the choice of the electrode material. The current ON-OFF ratio reaches a maximum value of $10^5$ and the mobility for both holes (p-channel) and electrons (n-channel) were 0.03 ×$10^{-3}$ cm²V⁻¹s⁻¹ and 0.78 ×$10^{-3}$ cm²V⁻¹s⁻¹ respectively. In the BGBC mode of operation, the BG changes both the carrier concentration of the N2200 channel and the $\Delta E_F$ of graphene for improving the electron and hole injection into the polymer.

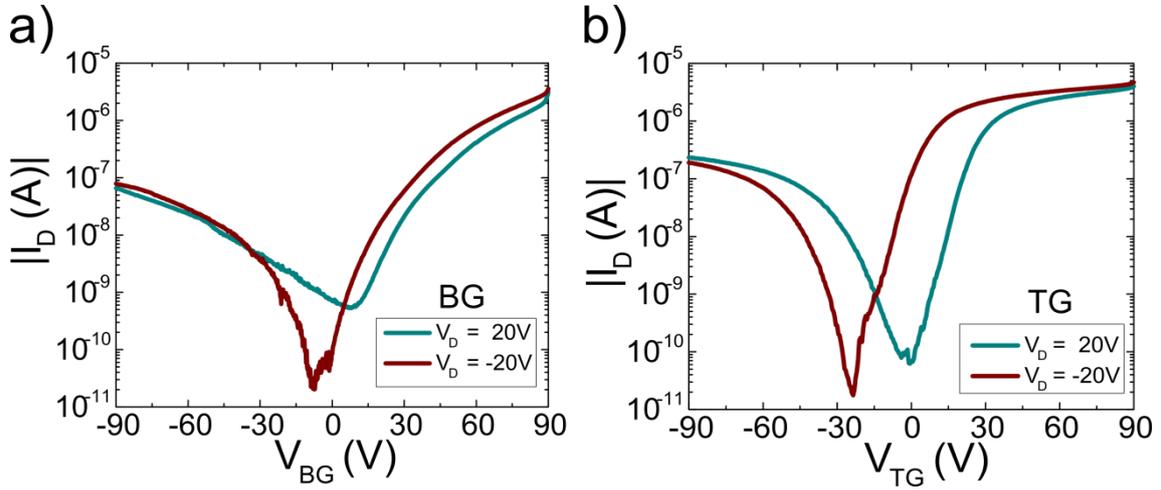

Figure 3: Ambipolar transfer characteristics of the N2200 transistor with (PMMA+CYTOP) top dielectric and with graphene as source and drain electrodes, a) probed with respect to bottom gate, b) probed with respect to top gate, for both the polarities of the drain bias $V_D$.

Figure 3b presents the transfer curve for the TGBC operation for both polarities of $V_D$. We observe a balanced ambipolarity similar to the one observed for the BGBC case, with a typical shift of the turn on voltage towards more negative value for a negative $V_D$ [26,28,31]. The current ON-OFF ratio was > $10^5$, while the mobility for both holes (p-channel) and electrons (n-channel) were 0.22 ×$10^{-3}$ cm$^2$V$^{-1}$s$^{-1}$ and 12.55 ×$10^{-3}$ cm$^2$V$^{-1}$s$^{-1}$, respectively. We observe that the performance of the device with graphene electrodes is better than that of Ti/Au electrode [28].

We can now explore the tuning of the N2200 ambipolarity. As presented above, the energy barrier for charge carrier injection into N2200 can be modulated due to the Fermi level tuning of graphene by the bottom gate bias. In the rigid band approximation, the $E_F$ of the graphene electrodes is at 4.65 eV from the vacuum level (considering the CNP of pristine CVD graphene at 4.5 eV from the vacuum level, see SI), while the LUMO and HOMO of N2200 are at 4.0 eV and 5.6 eV from the vacuum level, respectively (Figure 4a). In this approximation, the $E_F$ of graphene is at the middle of the band gap of N2200, a position which favors the injection of both electrons and holes into the semiconductor. The advantage for ambipolarity is clear in comparison with the work function of Au, which is positioned at 5.0 eV, and hence closer to the HOMO of the polymer. Furthermore, and specifically for graphene, $E_F$ can be moved upwards (downwards) to match well with the LUMO (HOMO) level of N2200 for a positive (negative) $V_{BG}$. The weak screening of the electric field from the bottom gate by the graphene electrodes also helps in the accumulation of extra electrons (holes) in the N2200 channel for positive (negative) $V_{BG}$.

Taking advantage of these two factors, the ambipolarity in N2200 can be modulated effectively by fixing the bottom gate bias (thereby fixing the energy level alignment and the induced charge carriers in the bottom channel) and varying the top gate bias ($V_{TG}$) (Figure 4d and 4e). A complete suppression of p-type transport (in the negative $V_{TG}$ regime) is possible for a high enough value of $V_{BG}$. Technically, for a given value of gate bias ($V_{BG}$ or $V_{TG}$), the accumulated charge in the N2200 is larger for BG than for TG due to the higher value of the BG capacitance ($C_{BG}$ = 11.5 nF.cm$^{-2}$ and $C_{TG}$ = 2.8 nF.cm$^{-2}$). However, because of the two counter-interacting electric fields and the accumulated charges, the N2200 channel is never fully depleted of charge carriers and hence a poor OFF state is observed in this case (Figure 4d and 4e). Similarly, by fixing $V_{BG}$ to negative values and sweeping $V_{TG}$, the electron current can be suppressed down to a certain extent as the N2200 polymer is intrinsically n-type (Figure 4f and 4g). The effective energy barrier modulation at the graphene/N2200 interface and the penetration of the bottom gate electric field through graphene are the main dominating factors in tuning the ambipolarity in N2200.

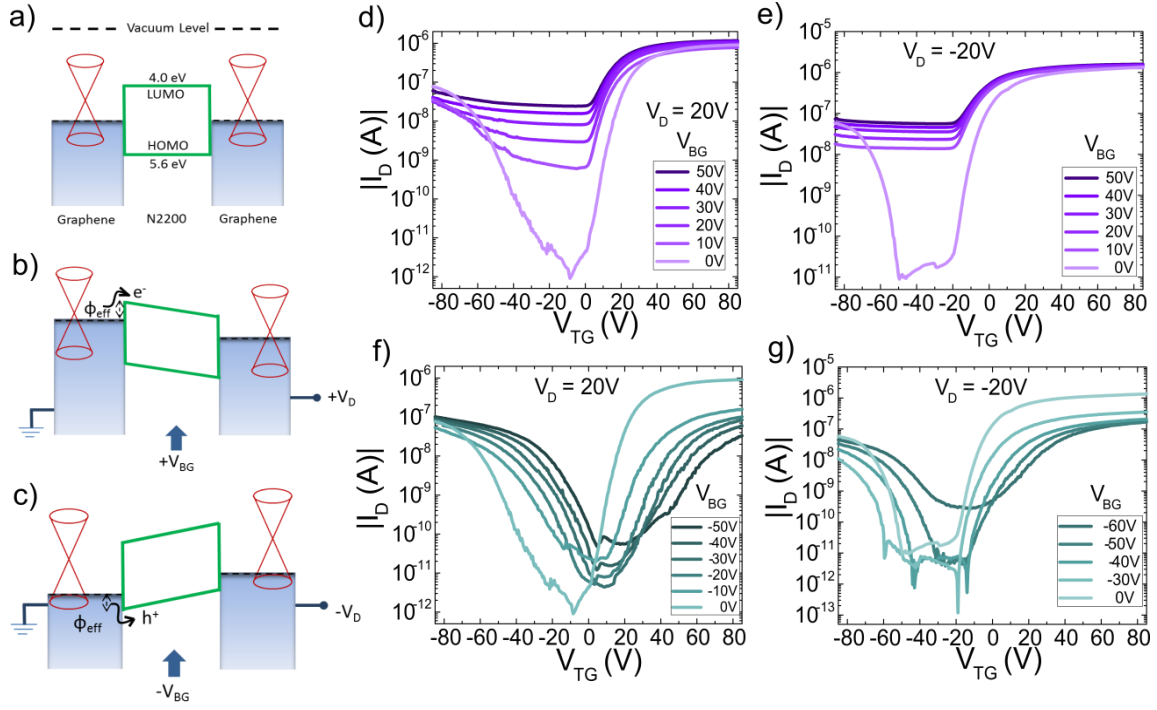

Figure 4: (a) Rigid band energy diagram of the graphene-N2200-graphene lateral transistor. The effective energy barrier for electron and hole injection into the semiconductor can be modulated by a positive $V_{BG}$ (b) and a negative $V_{BG}$ (c), respectively. Dual gating operation and suppression of the induced ambipolarity with fixed positive bottom gate voltages while varying the top gate bias with (d) $V_D = 20V$, (e) $V_D = -20V$. Increase of p-type and suppression of n-type conductivity of the devices with a fixed negative bottom gate voltage while varying the top gate bias with (f) $V_D = 20V$, (g) $V_D = -20V$.

In conclusion, we have demonstrated the successful operation of an ambipolar polymer FET with graphene electrodes. The n-type operation of the transistor with a pristine N2200 channel showed a current ON-OFF ratio of $>10^7$ and an electron mobility of $> 0.8 \times 10^{-3}$ cm$^2$V$^{-1}$s$^{-1}$, consistent with previous findings. Ambipolarity was induced in the N2200 polymer by the presence of a top polymer dielectric and could be detected from both BGBC and TGBC mode of operations using graphene electrodes, with better output characteristics than those reported for conventional metal electrodes. Moreover, our findings prove that the induced bulk ambipolarity in N2200 is independent of the choice of electrode materials. The effective energy barrier modulation and weak screening of the electric field of the graphene electrodes provided a way to tune the ambipolar response in a dual gating operation. Our results demonstrate the use of graphene electrodes as a practical route to tune and enhance the ambipolar characteristics of a polymer FET towards all carbon based large-area electronics.

**Experimental Section:**

***Transistor fabrication:*** We used $SiO_2$ (300 nm)/ $Si^{n++}$ as the substrate where the $SiO_2$ (300 nm) acted as the bottom gate dielectric and $Si^{n++}$ as the bottom gate contact. Large area CVD-grown graphene was transferred onto the $SiO_2$ (300 nm)/ $Si^{n++}$ substrate (as provided by Graphenea S.A) and individual chips of 1×1 $cm^2$ were used for the transistor fabrication. The graphene electrodes are defined with standard e-beam lithography (EBL) using double layer PMMA and etched through with Oxygen/Argon plasma. Ti/Au (3 nm/36 nm) contacts for the graphene bars are patterned with a second EBL step, followed by metal evaporation and lift-off. The samples were then vacuum ($< 10^{-7}$ mbar) annealed at 250˚C overnight to remove the residues from the lithography process. We used two 1500 µm long and 100 µm wide graphene bars on $SiO_2$ (300 nm)/$Si^{n++}$ for the bottom source (S) and the drain (D) electrodes. The distance between the graphene electrodes defines the channel length (L = 10 µm) and the length of the graphene bar defines the channel width (W = 1500 µm) with a W/L ratio of 150. The N2200 polymer (55 nm) was spin-coated on the sample and baked at 100˚C for 30 minutes. PMMA (60 nm) and CYTOP (630 nm) were then successively spin-coated on the transistor array to form the top gate dielectric. The sample was baked at 180°C for 2 minutes after PMMA spin-coating and 130°C for 1h 30 minutes after CYTOP spin-coating. Finally, a 25 nm Al layer was thermally evaporated on the transistor array through a shadow-mask to form the top gate contact.

*Thin films and transistor characterization:*

All the thicknesses of the solution processed films were pre-calibrated by spin-coating them on a $SiO_2$ (300 nm)/ $Si^{n++}$ substrate and measuring them with X-Ray Reflectivity (in an XPert PRO PANanalytical diffractometer). The thickness of the CYTOP film (630 nm) was characterized with a profilometer (Veeco Dektak150, Telstar Instruments). The electrical characterization of the graphene FET and the N2200 FET (before and after top dielectric deposition) was performed in a variable temperature Lakeshore probe station under high vacuum using a Keithley 4200-SCS semiconductor analyzer.

**Conflicts of interest**
There are no conflicts to declare.

**Acknowledgements**
This work is supported by the Spanish MICINN under the Maria de Maeztu Units of Excellence Programme (MDM-2016-0618) and under Project RTI2018-094861-B-100, and by the European Union H2020 under the Marie Curie Actions (796817-ARTEMIS and 766025-QuESTech).